\documentclass[pra,twocolumn, aps, 10pt, superscriptaddress]{revtex4-2}
\usepackage[english]{babel}
\usepackage{textcomp,xcolor,natbib}
\usepackage{graphicx}
\usepackage{amsmath, bbm}
\usepackage{latexsym}
\usepackage{amsfonts}   
\usepackage{amssymb}
\usepackage{array}      
\usepackage{epsfig}
\usepackage{txfonts}
\usepackage{xcolor,braket}
\usepackage[normalem]{ulem}
\usepackage{soul}
\usepackage{dsfont}
\usepackage{xfrac}  
\usepackage{relsize}
\usepackage{enumitem}
\usepackage{float}
\usepackage{dcolumn}
\usepackage{bm}

\begin{document}

\title{Control of memory effects in a spin-boson system by periodic driving}

\author{Pietro Marco Follia}
\affiliation{Dipartimento di Fisica ``Aldo Pontremoli", Università degli Studi di Milano, via Celoria 16, 20133 Milan, Italy}
\affiliation{Istituto Nazionale di Fisica Nucleare, Sezione di Milano, via Celoria 16, 20133 Milan, Italy}
\author{Bassano Vacchini}
\affiliation{Dipartimento di Fisica ``Aldo Pontremoli", Università degli Studi di Milano, via Celoria 16, 20133 Milan, Italy}
\affiliation{Istituto Nazionale di Fisica Nucleare, Sezione di Milano, via Celoria 16, 20133 Milan, Italy}

\author{Heinz-Peter Breuer}
\affiliation{Institute of Physics, University of Freiburg, 
Hermann-Herder-Stra{\ss}e 3, D-79104 Freiburg, Germany}
\affiliation{EUCOR Centre for Quantum Science and Quantum Computing,
University of Freiburg, Hermann-Herder-Stra{\ss}e 3, D-79104 Freiburg, Germany}

\date{\today}

\begin{abstract}
We study the emergence of quantum memory effects in a spin-boson system at finite 
temperature driven by an external time-periodic force. Quantifying memory effects
by the trace-distance based measure for non-Markovianity and performing numerical
simulations employing the hierarchical equations of motion approach, we find a
pronounced peak structure when plotting the non-Markovianity measure
as a function of the driving amplitude. This distinctive feature is interpreted
using Floquet theory and the Floquet-Lindblad master equation, associating the peaks with 
the degeneracies of the quasienergy spectrum which lead to a strong enhancement of the 
relaxation times of the system. These results suggest strategies for the efficient control 
of non-Markovianity in open quantum systems by periodic driving.
\end{abstract}

\maketitle

\section{Introduction}
Any realistic quantum system is almost never perfectly isolated: it rather interacts with its surroundings, exchanging both energy and information. Such systems are naturally described within the framework of open quantum systems~\cite{Breuer_book, Vacchini2024}. The interaction with a reservoir typically leads to dissipation and decoherence. Beyond these well-understood effects, the environment can also act as a memory reservoir, giving rise to memory effects, i.e. non-Markovian dynamics. Non-Markovianity can qualitatively alter the behavior of quantum systems, affecting coherence times and entanglement dynamics~\cite{Rivas2014a,Breuer2016a,Li2018a}. Understanding these effects is therefore of fundamental importance, both for foundational aspects of quantum theory and for practical applications in quantum technologies.

A paradigmatic model for investigating open-system dynamics is the spin-boson model~\cite{Rev_Legget}. Despite its apparent simplicity, it captures key features of a wide range of phenomena, including electronic transfer~\cite{SB_Elect_transf} and macroscopic quantum coherence~\cite{SB_macro_coher}. Of particular relevance to recent efforts aimed at protecting quantum coherence via external control is the periodically driven spin-boson model~\cite{one_over_f_noise,Exp_protection_drive_super}.
While non-Markovianity in the undriven spin-boson model has been extensively studied~\cite{NM_sb_any,NM_Sb_zero}, its role in periodically driven variants remains largely unexplored, and has received attention only recently~\cite{amati_review}.

In this work, we employ the non-Markovianity measure introduced in~\cite{non_markov_orig} to investigate the impact of periodic driving on memory effects in the dynamics of the open system (spin) in the spin-boson system.
To this end, we have carried out extensive numerical simulations of the driven spin-boson model using the hierarchical equations of motion (HEOM) approach~\cite{Tanimura2020a}.
Our results uncover a distinctive and unexpected peak-like structure in the non-Markovianity of the system as a function of the driving amplitude. We further observe
a similar structure in the relaxation times of the open system with peaks which are in one-to-one correspondence with the peaks
of non-Markovianity. We demonstrate that these results can be understood theoretically by utilizing the quasienergy spectrum and the
Floquet-Lindblad master equation~\cite{Blumel1991,Breuer1997,Breuer_book} of the system.
Indeed, each peak corresponds to a degeneracy of the quasienergies, leading to a significant change of the structure of the Lindblad jump operators of the master equation. 
In the degenerate case only two dissipative channels remain active, compared to three channels 
outside the resonances. This yields a strong increase of the relaxation times of the spin 
leading to the emergence of an almost decoherence-free subspace and
long-lived information backflow processes resulting in peaks of the non-Markovianity.
This physical mechanism paves the way for the development of strategies to control relaxation 
rates and memory effects in open systems through periodic driving of the open system.
The manuscript is organized as follows. In Sec.~\ref{nm} we introduce the definition of non-Markovianity used in this analysis and explain how it is quantified. In Sec.~\ref{dsb} we present the physical model of driven open system considered in this work. In Sec.~\ref{flms} we briefly review the Floquet formalism relevant for describing this model within the Markovian semigroup approximation. In Sec.~\ref{sec_num_resu_theo_exp} we present our numerical results and discuss their physical interpretation. Finally, in Sec.~\ref{cao}, we summarize and comment on our findings.

\section{Non-Markovianity}\label{nm}
Defining non-Markovian processes in quantum theory is a non-trivial task, as the concepts employed in classical probability theory do not naively generalize to the quantum setting~\cite{non_markov_2011, non_markov_2015,Breuer2016a,Chruscinski2022a}. In recent years, several approaches have been proposed to define and measure non-Markovianity for open quantum systems~\cite{non_markov_orig,non_markov_entang, non_markov_corr}. In this work we employ the definition developed in~\cite{non_markov_orig}, which is based on the notion of information backflow.
The central quantity of this approach is the distinguishability between two quantum states of the reduced system $\rho^A$ and  $\rho^B$ quantified by means of the trace distance~\cite{QInfo,QCmp_QInfo}:
    \begin{equation}
        D(\rho^A, \rho^B) = \frac{1}{2}\text{Tr} \, \big| \rho^A-\rho^B \big|,
    \end{equation}
    where $|A| = \sqrt{A^\dagger A}$. 
    If the dynamics of the open system is described by the family 
    $\{ \Phi_t \mid t\geq 0 \}$ of quantum dynamical maps $\Phi_t$, then 
    for any pair of initial states $\{\rho^A,\rho^B\}$ the time evolution of the trace distance is given by
    \begin{equation}
        D_t\big(\rho^A, \rho^B\big)  = D\big(\Phi_t[\rho^A], \Phi_t[\rho^B]\big).
    \end{equation}
    A monotonically decreasing trace distance, i.e. $ \dot {D}_t \leq 0$, implies that the two states are becoming increasingly less distinguishable, which is interpreted as a unidirectional flow of information from the system to the environment. This clearly defines a Markovian dynamics as the information initially contained in the reduced system is lost in the environment and never retrieved. Conversely, an increase of the trace distance over time, i.e. $\dot{D}_t > 0$, is interpreted as information flowing from the environment into the system. This phenomenon, known as information backflow, is regarded as the characteristic feature of non-Markovian dynamics~\cite{non_markov_orig,Megier2021a}.
Based on this interpretation, a quantitative measure of non-Markovianity for the 
process described by the maps $\Phi_t$
is defined by
    \begin{equation}
    \label{eq:NM}
        \mathcal{N}[\Phi] = \max_{\rho^A,\rho^B} \int_{\dot{D}_t > 0} \!\!\! dt \, \dot{D}_t ,
    \end{equation}
    where the maximum is taken over all possible orthogonal pairs of initial states.

\section{Driven spin-boson model}\label{dsb}
The spin-boson model is a paradigmatic model in the theory of open quantum systems providing a perfect venue to understand dissipative effects~\cite{Weiss_book,Breuer_book, Rev_Legget}. It describes a two level system with Hamiltonian $({\omega_0}/{2})\sigma_z$ coupled to a bath of harmonic oscillators with Hamiltonian $H_B = \sum_n\omega_n a^{\dagger}_n a_n$ by a linear interaction term
    \begin{equation}
       H_{I} = \sigma_x\otimes\sum_n g_n(a^\dagger_n+a_n),
    \end{equation}
    where $\sigma_z = \ket{e}\bra{e}-\ket{g}\bra{g}$, $\sigma_x= \ket{e}\bra{g}+\ket{g}\bra{e}$ and $a_n^\dagger$, $a_n$ denote the bosonic creation and annihilation operators associated with the harmonic oscillator of frequency $\omega_n$. Finally, the coefficients $g_n$ describe the distribution of the couplings between the system and the different harmonic modes.

    In this work, we consider in particular a variation of this model that includes a driving term modeling, e.~g. the interaction with a classical monochromatic radiation field on resonance, so that the complete system Hamiltonian reads
    \begin{equation}
    \label{eq:HD}
        H_S(t) = \frac{\omega_0}{2}\sigma_z-\Omega \cos(\omega_0 t)\sigma_x,
    \end{equation}
where $\Omega$ denotes the driving strength and we use units such that $\hbar=1$.
    The influence of the environment on the open system can be expressed using the spectral density~\cite{Weiss_book,Breuer_book,Vacchini2024},
    which encodes the information on couplings and frequencies distribution relevant for the reduced dynamics. We assume a spectral density of the Lorentz-Drude form,
    \begin{equation}
    \label{LD_Spectral}
        J(\omega) = \alpha\omega\frac{\omega_c}{\omega^2+\omega_c^2},
    \end{equation}
    where $\omega_c$ represents the cutoff frequency and $\alpha$ the coupling strength between system and environment.
    This widely used expression is particularly suitable for the 
    numerically exact HEOM approach~\cite{Tanimura2020a,Johansson2013a,Julia_HEOM} employed in the present work.
    
    We further assume an initial product state of system and environment, 
    $\rho_{SB}(0) = \rho_S(0)\otimes\rho_B$
    with the bath in the thermal state $\rho_B = e^{-\beta H_B}/\mathcal{Z}$.
    With these assumptions, a formally exact expression for the dynamics of the reduced system can be written in the path-integral formalism, where all the information regarding the environment is encoded in the correlation function~\cite{FVInfluence}:
    \begin{equation}
        C(t) = \sum_n|g_n|^2\langle (a_n(t)+a^\dagger_n(t))(a_n(0  )+a^\dagger_n(0))\rangle_{\rho_B},
    \end{equation}
which can be expressed in terms of the spectral density of the environment according to
    \begin{equation}
    \label{eq:Fluct_diss_theo}
        C(t) = \frac{1}{2\pi}\int d\omega e^{-i\omega t}\frac{J(\omega)}{1-e^{-\beta \omega}}.
    \end{equation}
      Within the HEOM approach the integral \eqref{eq:Fluct_diss_theo} can be computed by means of the residue theorem in the form $C(t) = \sum_l \eta_le^{-\gamma_lt}$,
    where $\{\eta_l,\gamma_l\}$ are determined according to the Padé decomposition scheme~\cite{Pade_Optimal},
    and the reduced system density operator is obtained resorting to a hierarchy of coupled equations for auxiliary density operators, which can be truncated and solved numerically with high accuracy. This provides access to the full reduced dynamics, enabling a direct computation of the non-Markovianity measure as defined in Eq.~\eqref{eq:NM}.

\section{Floquet-Lindblad master equation}\label{flms}
    The periodicity of the driving allows to exploit the extension of Floquet theory to open quantum systems.
    Floquet's theorem states that the time-evolution operator determined by a time-periodic Hamiltonian, $H_S(t) = H_S(t+T)$ with $T=2\pi/\omega_0$, admits the representation
    \begin{equation}
    \label{eq:floquet_theorem}
        U(t) = \sum_k \ket{u_k(t)} \bra{u_k(0)} e^{-i\epsilon_k t},
    \end{equation}
    where the states $\ket{u_k(t)}$ are known as Floquet states, which form an orthonormal basis of the Hilbert space for each fixed time $t$, and the real constants 
        $\epsilon_k$ are the associated quasienergies~\cite{Breuer1988a,Breuer1988b,Zeldovich1967a,Shirley1965a,Ritus1967a,Sambe1973a,Goldman2014a,Holthaus2015a}. 
    This decomposition offers a natural framework for deriving in the weak-coupling limit and for not too low driving strength a time-dependent Markovian master equation for periodically driven open systems, known as Floquet-Lindblad master equation~\cite{Blumel1991,Breuer1997,Breuer_book,Ikeda2021a,Mori2023a}, which we will use to benchmark and theoretically explain our results obtained by means of numerically exact techniques. In the interaction picture it takes the form 
    \begin{equation} \label{MEQ}
    \frac{d}{dt} \rho_S(t) = \mathcal{D}[\rho_S(t)]
    \end{equation}
    where the dissipator is given by
\begin{equation}
\label{eq:Dissipa}
    \mathcal{D}[\rho] = \sum_{\omega_F} \gamma(\omega_F)\left[A(\omega_F)\rho A^\dagger(\omega_F) - \frac{1}{2}\{A^\dagger(\omega_F)A(\omega_F), \rho\}\right],
\end{equation}
with jump operators expressed in the Floquet basis as
\begin{equation}
\label{eq:A_omega_f}
    A(\omega_F) = 
    \sum_{i,j,n} 
    c^n_{ij} \ket{u_i(0)}\bra{u_j(0)},
\end{equation}
and decay rates
\begin{equation}    
    \gamma(\omega_F) = \frac{J(\omega_F)}{1 - e^{-\beta \omega_F}}.
\end{equation}
The Floquet frequencies $\omega_F$ are given by differences between quasienergies and harmonics of the driving frequency, so that the sum in \eqref{eq:A_omega_f} is to be extended over the set of indexes $i,j,n$ that satisfy the condition $\epsilon_j-\epsilon_i-n\omega_0 = \omega_F$. The coefficients are the Fourier components of the coupling operator in the Floquet basis, namely
\begin{equation}
    c^n_{ij} = \int_0^T \frac{dt}{T} e^{-i n \omega_0 t} \bra{u_i(t)} \sigma_x \ket{u_j(t)}.
    \label{eq:four_coef}
\end{equation}

\section{Numerical results and theoretical interpretation}
\label{sec_num_resu_theo_exp}
    In the simulation, the environmental parameters were kept fixed while increasing 
 the driving amplitude  $\Omega$, and the non-Markovianity measure $\mathcal{N}$ of Eq.~\eqref{eq:NM} was computed by maximizing over a thousand randomly drawn initial pairs of orthogonal states.
    \begin{figure}[htb]
        \centering
        \includegraphics[width=\columnwidth]{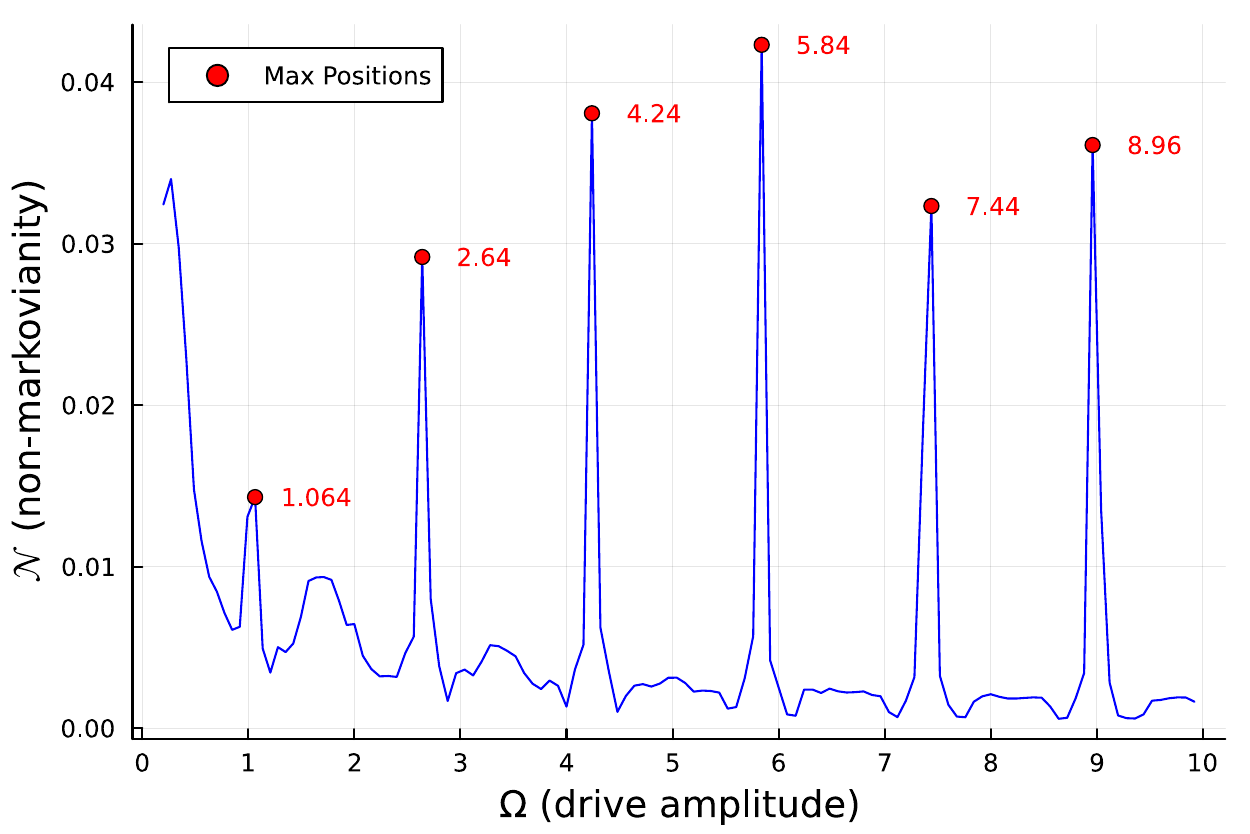}
        \includegraphics[width=\columnwidth]{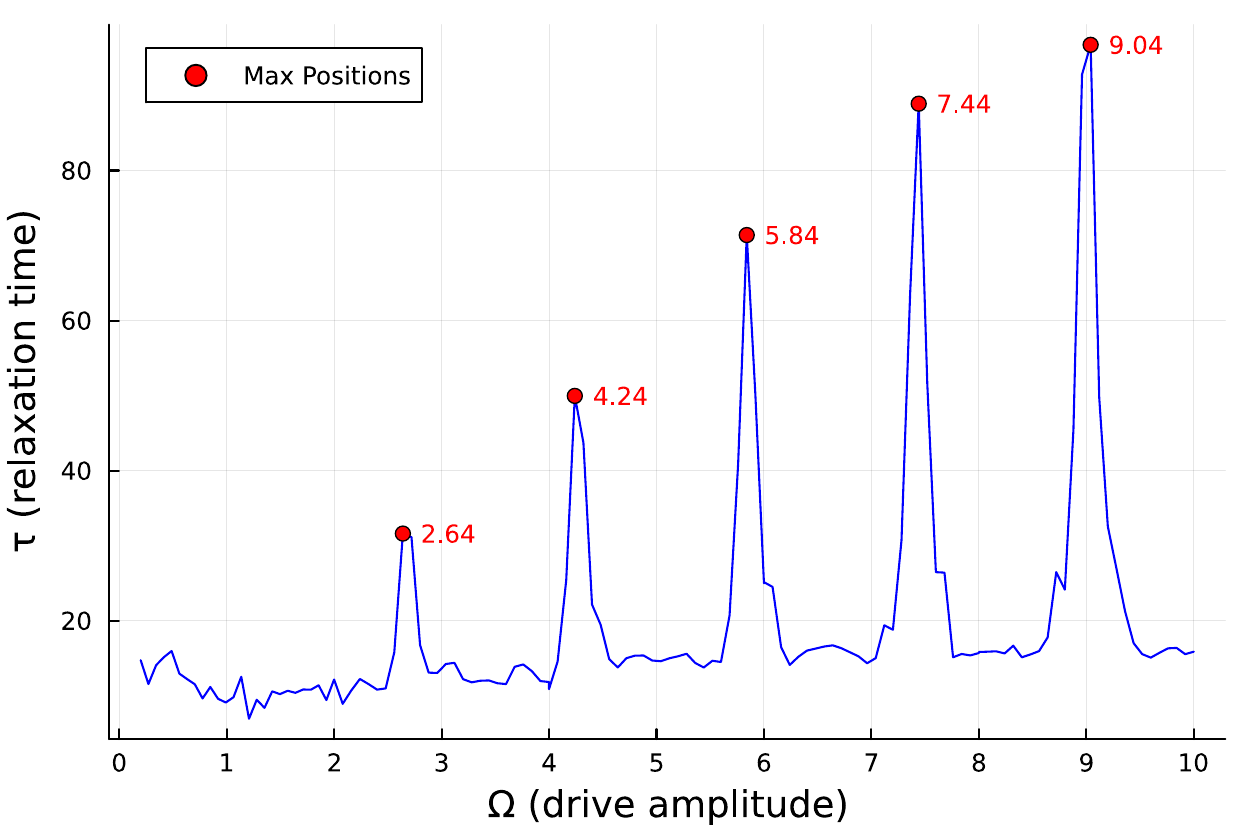}
        \caption{Non-Markovianity (top) and relaxation time (bottom) of the driven spin-boson model as a function of the driving amplitude $\Omega$ for resonant driving. Parameters in units of the system frequency $\omega_0$ are $k_BT = 1.0$, $\alpha = 0.1$ and $\omega_c = 1.0$.
        }
        \label{Fig:non_markovyrela_times}
    \end{figure}

\begin{figure}[htb]
        \centering
        \includegraphics[width=\columnwidth]{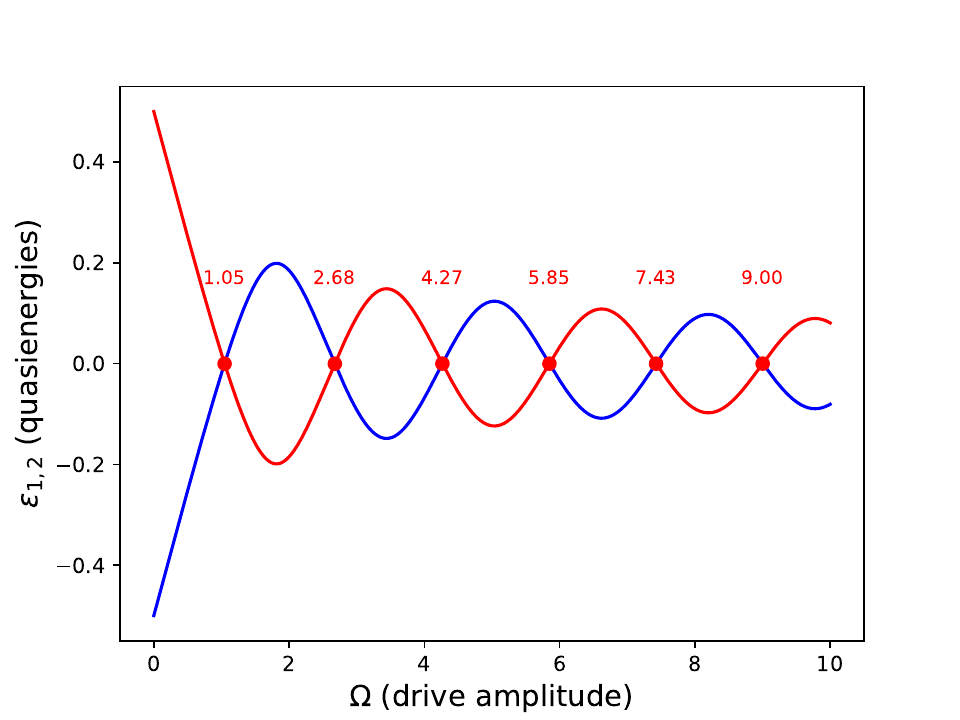}
        \caption{Quasienergies of the driven spin system as a function of the driving amplitude $\Omega$, for the same parameters considered in Fig.~\ref{Fig:non_markovyrela_times}.}
        \label{Fig:quasi}
    \end{figure}

As shown in Fig.~\ref{Fig:non_markovyrela_times}~(top) the results exhibit a distinctive 
peak-like structure. In order to explain this behavior, we carried out a systematic numerical investigation of the relaxation rates as a function of the driving amplitude $\Omega$. The time evolution of the matrix elements $\rho^S_{ij}(t)$ of the reduced system density matrix was fitted to an exponentially decaying envelop function of the form $f_{ij}(t) \propto e^{-t/\tau_{ij}}$ in order to extract the decay time $\tau_{ij}$ of each element. The overall system decay time was then defined as 
$\tau = \max_{i,j} \tau_{i,j}$. The extracted decay times $\tau$ provide a quantitative estimate of the relaxation rates and are depicted in Fig.~\ref{Fig:non_markovyrela_times}~(bottom).
    The simulations reveal a peak-like structure also for the relaxation time, with a clear correspondence to the non-Markovianity peaks. We can gain further physical insight into the origin of these numerically observed effects by analyzing the dependence of the system’s quasienergies appearing in Eq.~\eqref{eq:floquet_theorem} on the driving amplitude $\Omega$.
    These quasienergies can be computed numerically using standard tools~\cite{Johansson2013a} and are shown in Fig.~\ref{Fig:quasi}. Upon comparison with Fig.~\ref{Fig:non_markovyrela_times} we observe a clear one-to-one correspondence between quasienergy crossings and peaks in non-Markovianity, suggesting a deeper physical origin for the peak structure.
    
    To provide a microscopic understanding of the obtained results we consider the Floquet-Lindblad master equation Eq.~\eqref{MEQ}, which provides a good approximation for the dynamical description in the weak-coupling regime, in both the degenerate case in which the quasienergies coincide,  and the non-degenerate regimes away from the quasienergy crossings.
    
Starting from the non-degenerate case, namely $\epsilon_1 \not = \epsilon_2$, we determine the jump operators according to Eq.~\eqref{eq:A_omega_f} by numerical evaluation of the Fourier components defined in Eq.~\eqref{eq:four_coef},
which yields $c^n_{1,1} = -c^n_{2,2}$, $|c^n_{1,1}|=|c^n_{2,2}| \approx  \tfrac{1}{5}(\delta_{n,1}+\delta_{n,-1})$ and $c^n_{1,2}=(c^n_{2,1})^* \approx \delta_{n,0}$
(see Appendix \ref{eval_four_coeff} for further details).
The Lindblad jump operators for the non-degenerate case are thus given by
        \begin{align}
        \label{jump_eff}
         A(\epsilon_2 -\epsilon_1) &=  \ket{u_1}\bra{u_2} \equiv \Sigma_+\nonumber \\
        A(\omega_0) &= \tfrac{1}{5}(\ket{u_1}\bra{u_1} -\ket{u_2}\bra{u_2}) \equiv \tfrac{1}{5}\Sigma_z ,
        \end{align}
where $\ket{u_j}\equiv\ket{u_j(0)}$ was used to simplify notation, and the Floquet
basis has been used to define a representation of the $\mathfrak{su}(2)$ algebra
with Pauli operators $\Sigma_{\pm}$ and $\Sigma_z$. Thus, we have three Lindblad
operators: The operators $A(\omega_F)$ and $A^\dagger(\omega_F)=A(-\omega_F)$
describe upward and downward transitions between the Floquet states $\ket{u_1}$
and $\ket{u_2}$ with transition frequency
$\omega_F=\pm(\epsilon_1-\epsilon_2)$, while $A(\omega_0)$ represents elastic scattering
processes at the resonant frequency $\omega_F=\omega_0$.
        According to Eq.~\eqref{eq:Dissipa}, given the jump operators defined in Eq.~\eqref{jump_eff} the Floquet-Lindblad dissipator in the master equation for the non-degenerate case is given by:
    \begin{align}
    \label{eq:FL1}
      \mathcal{D}_{\mathrm{ND}}[\rho_S(t)]
        =&\gamma_{\downarrow}[\Sigma_-\rho_S(t)\Sigma_+-\tfrac{1}{2}\{\Sigma_+\Sigma_-, \rho_S(t)\}]\nonumber \\
        &+\gamma_{\uparrow}[\Sigma_+\rho_S(t)\Sigma_--\tfrac{1}{2}\{\Sigma_-\Sigma_+, \rho_S(t)\}] \nonumber\\
        &+\frac{\gamma_{z}}{25}[\Sigma_z\rho_S(t)\Sigma_z-\rho_S(t)],
    \end{align}
    where $\mathrm{ND}$ stands for non-degenerate and the rates read  
    \begin{align}
      \gamma_{\downarrow} &=J(\epsilon_2 -\epsilon_1)(N(\epsilon_2 -\epsilon_1)+1),
        \quad
      \gamma_{\uparrow}=J(\epsilon_2 -\epsilon_1)N(\epsilon_2 -\epsilon_1)
       \nonumber \\
      \gamma_{z} &=J(\omega_0)(2N(\omega_0)+1),
    \end{align}
with $N(\omega) = (e^{\beta\omega}-1)^{-1}$  the Bose-Einstein distribution.
    
    We can now solve for the time evolved density matrix of the reduced system using the basis defined by the Floquet states, namely $\rho_S^{i,j}(t)  = \bra{u_i}\rho_S(t)\ket{u_j}$.
    The populations $\rho_S^{i,i}(t)$ decay exponentially to their equilibrium values with relaxation time $\tau_{\mathrm{ND}}^{\mathrm{diag}}=1/(\gamma_{\downarrow}+\gamma_{\uparrow})$,
    while the off-diagonal elements $\rho_S^{i,j}(t)$ are exponentially suppressed on the time-scale  $\tau_{\mathrm{ND}}^{\mathrm{off-diag}}=1/{[(\gamma_{\downarrow}+\gamma_{\uparrow})/2+2\gamma_{z}/25]}$.
  
    The degenerate case, in which the two quasienergies of the system coincide, so that $\epsilon_1 =\epsilon_2$ and a crossing occurs, leads to a different set of jump operators. Indeed, exploiting the previous evaluation of the Fourier components we find
    \begin{align} \label{jump_eff-deg}
    A(0) &=\ket{u_2}\bra{u_1} + \ket{u_1}\bra{u_2} \equiv \Sigma_x \nonumber \\
    A(\omega_0) &= \frac{1}{5}(\ket{u_1}\bra{u_1}-\ket{u_2}\bra{u_2}) \equiv \frac{1}{5}\Sigma_z,
    \end{align}
    which yields the Floquet-Lindblad dissipator
    \begin{align}
    \label{eq:FL2}
      \mathcal{D}_{\mathrm{D}}[\rho_S(t)]
        = \gamma_{x} [\Sigma_x\rho_S(t)\Sigma_x-\rho_S(t)]
         +\frac{\gamma_{z}}{25}[\Sigma_z\rho_S(t)\Sigma_z-\rho_S(t)],
    \end{align}
    where  $\mathrm{D}$ stands for degenerate and we have further introduced the rate
    \begin{align} \label{gamma_x}
      \gamma_{x} &=\lim_{\omega\to 0^+}J(\omega)(2N(\omega)+1).
    \end{align}
While the jump operator $A(\omega_0)$ for elastic scattering and the associated
rate $\gamma_z$ are the same as in the non-degenerate case, transitions between
Floquet states $\ket{u_1}$ and $\ket{u_2}$ 
are described by the jump operator $A(0)=\Sigma_x$ with transition rate $\gamma_x$ given by \eqref{gamma_x}. The emergence of the jump operator $\Sigma_x$ can be understood from the
fact that both the transition $\ket{u_1} \to \ket{u_2}$ and the reverse
transition $\ket{u_2} \to \ket{u_1}$ correspond to zero transition frequency
and, hence, the associated jump operators add coherently.

    Similarly to the non-degenerate case, the dynamics can be solved using the Floquet-state basis. For the diagonal elements, we still have exponential relaxation to the equilibrium value with decay time $\tau_{\mathrm{D}}^{\mathrm{diag}}=1/\gamma_{x}$. 
However, the real and imaginary part of the coherences, while still exponentially vanishing, have different relaxation times, namely        
\begin{align}
\label{eq:larger}
      \tau_{\mathrm{D}}^{\mathrm{off-diag}}(\Re) &=25/(2\gamma_{z})\nonumber \\
        \tau_{\mathrm{D}}^{\mathrm{off-diag}}(\Im) &=1/[2(\gamma_{x}+\gamma_{z}/25)],
\end{align}
comparing Eq.~\eqref{eq:FL1} and Eq.~\eqref{eq:FL2}, we can immediately observe a fundamental change in the dissipation channels  appearing in the Floquet-Lindblad master equation. Indeed, in the non-degenerate case of Eq.~\eqref{eq:FL1} three jump operators appear, namely $\Sigma_{+}$, $\Sigma_{-}$ and $\Sigma_z$, whereas in the degenerate case, only two dissipation channels are present: $\Sigma_x$ and $\Sigma_z$.
Importantly, since all $\gamma$'s are of the same order of magnitude, although all matrix elements decay exponentially in both cases, the different combination of jump operators leads to the fact that the relaxation time associated with the real part of the coherences in the degenerate case is significantly longer than all other time-scales:
\begin{equation}
    \tau_{\mathrm{D}}^{\mathrm{off-diag}}(\Re) \gg \tau_{\mathrm{D}}^{\mathrm{off-diag}}(\Im),\tau_{\mathrm{ND}}^{\mathrm{off-diag}},\tau_{\mathrm{D}}^{\mathrm{diag}},\tau_{\mathrm{ND}}^{\mathrm{diag}},
\end{equation}
as follows from Eq.~\eqref{eq:larger}.
This feature highlights the emergence of a nearly decoherence-free subspace along the direction of the $x$ component in the Floquet basis. In such a way, we can explain the longer relaxation times observed numerically at the quasienergies crossings. We further note that, in our analysis, the role of the spectral density is limited to the determination of the relative weights of the decay coefficients, which makes the presented mechanism  of rather general nature. These longer relaxation times allow for more information backflow processes to take place resulting in the non-Markovianity peaks observed numerically. Indeed, as shown in Appendix \ref{Appendix_trace_dist_behav}, the non-Markovian behavior is determined by small but persistent information backflow processes that accumulate during the dynamics. The longer the relaxation time, the higher the non-Markovianity measure of Eq.~\eqref{eq:NM}. Since this backflow process may persist until the last component of the reduced system reaches its limit value, it is ultimately determined by the maximum among the relaxation times, thereby justifying the choice presented in Fig.~\ref{Fig:non_markovyrela_times}. Indeed, as discussed below, the optimal pair maximizing the measure of Eq.~\eqref{eq:NM} corresponds to eigenstates of $\Sigma_x$, whose coherences decay on a timescale given by $\tau_{\mathrm{D}}^{\mathrm{off-diag}}(\Re)$.
This analysis further accounts for the monotonic increase with driving amplitude of the relaxation time peaks' height that can be observed in Fig.~\ref{Fig:non_markovyrela_times}. This behavior stems from the slight decrease of the value of the coefficient $c^1_{1,1}$ [Eq.~\eqref{eq:four_coef}] for increasing $\Omega$ when evaluated in correspondence of the quasienergies crossings (see Appendix \ref{eval_four_coeff} for details), which leads to a corresponding increase in $\tau_{\mathrm{D}}^{\mathrm{off-diag}}(\Re)$. Note that this argument holds in the regime of validity of the Floquet-Lindblad master equation, that is, for driving strength high enough, which explains the missing first peak in the relaxation times of Fig.~\ref{Fig:non_markovyrela_times}. In fact, for the corresponding value of $\Omega$ neither the value of $\Re(c^1_{1,2})$ is close to one, nor can $|c^1_{1,1}|$ be approximated by $\tfrac{1}{5}$.

We stress that this mechanism of suppression of decoherence
does not depend on a specific choice of parameters and is remarkably robust. In Appendix \ref{Appendix_other_params} we analyze the behavior of relaxation times and non-Markovianity for non-resonant driving and different cutoff frequency, that confirms the appearance of the peaks and their correlation with the quasienergy crossings, further demonstrating the reliability of our findings.
Our analysis differs substantially from 
previous suggestions in the literature. In~\cite{hanggi2004a} decoherence 
suppression was observed at quasienergy crossings but on the basis of a different mechanism
without emergence of a decoherence-free subspace. The study was conducted relying on high-frequency and static approximations to account for the influence of the driving, so that it requires very low temperatures. In the present work, we treat the Floquet–Lindblad master equation numerically evaluating the corresponding Fourier coefficients, so as to obtain a finer analysis that puts into evidence the appearance of a protected channel that allows for peaks of the non-Markovianity measure that are determined by the presence of a suitable protected pair of states.
The protection mechanism considered in~\cite{one_over_f_noise} appeared at extremal points of the
quasienergies and considered specific sources of noise, it is no longer valid in the presence of quasienergies crossings. In the recent work~\cite{Blais2025}, the authors identify a coherence–protection mechanism that is again linked to the presence of quasienergy degeneracies. Their analysis, however, relies on a partial secular approximation that does not capture the subtle interplay between the degeneracies and the behaviour of the exact Fourier coefficients $c^n_{ij}$. The latter govern the existence of a decoherence–free channel in the time evolution, a feature that is crucial for the emergence of the non-Markovianity peaks.

\begin{figure}[htb]%
    \centering
    \includegraphics[width=\columnwidth]{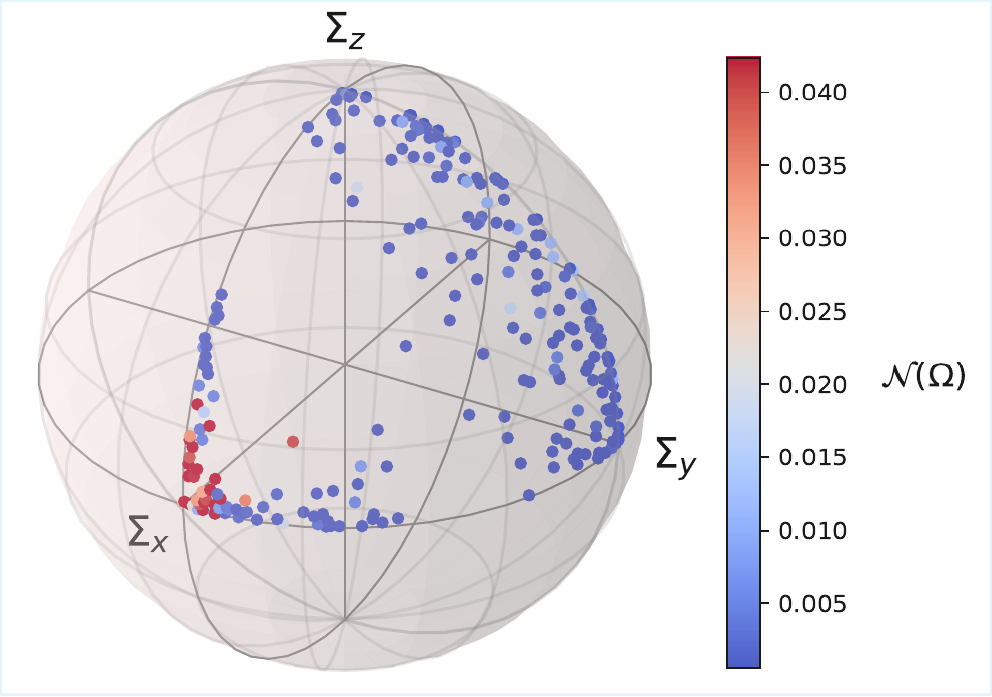}
    \caption{Pure states defining optimal pairs which maximize the non-Markovianity measure $\mathcal{N}$ for each value of the driving amplitude $\Omega$. To simplify the visualization, all states are displayed in the positive octant of the Bloch sphere, with the corresponding non-Markovianity measure represented as a heatmap. The highest value of non-Markovianity corresponding to the red points are indeed obtained for pure states in the $x$ direction.}
    \label{fig:heat_map_non_markov}
\end{figure}

The interpretation given above is further supported by the detailed investigation of the initial pair of states maximizing the non-Markovianity measure of Eq.~\eqref{eq:NM}, which according to~\cite{Wissmann2012a} has to be orthogonal and therefore given by pure states for a two-level system. Each pair is thus determined by a single pure state. We have numerically determined the maximizing pairs of pure states for different values of the driving amplitude $\Omega$ and visualized the results in Fig.~\ref{fig:heat_map_non_markov}. It clearly appears that the states leading to the highest value of non-Markovianity are given by the eigenstates of $\Sigma_x$, namely in Bloch notation by the states
   $\rho^{A,B}_S(0) = \frac{1}{2}(\mathbb{I} \pm \Sigma_x)$.
Indeed, this pair of states maximizes the real part of the coherence which is protected from dissipative effects in the degenerate case, thus leading to long-lasting processes of information backflow.

\section{Conclusions and outlooks}\label{cao}
We have investigated the role of memory effects in the spin-boson system under external periodic driving, highlighting and explaining a peak structure in the non-Markovianity measure based on the trace distance approach~\cite{non_markov_orig} as a function of the driving strength. The numerical evidence obtained using the HEOM technique has been interpreted by means of Floquet theory, through the analysis of the operator structure of the Floquet-Lindblad master equation. At driving strengths corresponding to crossings of the Floquet quasienergies, the master equation undergoes a structural change and a nearly decoherence-protected subspace arises. This is evidenced by the emergence of much longer relaxation times and a corresponding signature in the non-Markovianity measure. These findings provide complementary insights into coherence protection compared to previous work~\cite{hanggi2004a,one_over_f_noise, Blais2025}, suggesting a universally valid mechanism that does not rely on low-temperature regimes or highly specific environmental features. 
Furthermore, our results reveal a peculiar mechanism linking extended coherence times and enhanced non-Markovianity, to the extent that the maximizing pair of initial states can be identified at the quasienergy crossings by the one maximizing the coherence time, suggesting a strategy for the efficient control of memory effects by periodic driving.
This opens new avenues for applications requiring long-time coherence preservation in open quantum system dynamics, such as solid-state qubits used for quantum computation.
\section*{Acknowledgments}
B.V. acknowledges support by the Italian Ministry of Research and Next Generation EU via the NQSTI-Spoke1-BaC project QSynKrono (contract n. PE00000023-QuSynKrono).

\appendix

\section{Trace distance behavior}
\label{Appendix_trace_dist_behav}
In this Appendix we provide evidence for the attribution of the non-Markovianity peaks to the appearance of longer coherence times. To this aim we analyze the time evolution of the trace distance. 
\begin{figure}[htb]
    \centering
        \includegraphics[width=\columnwidth]{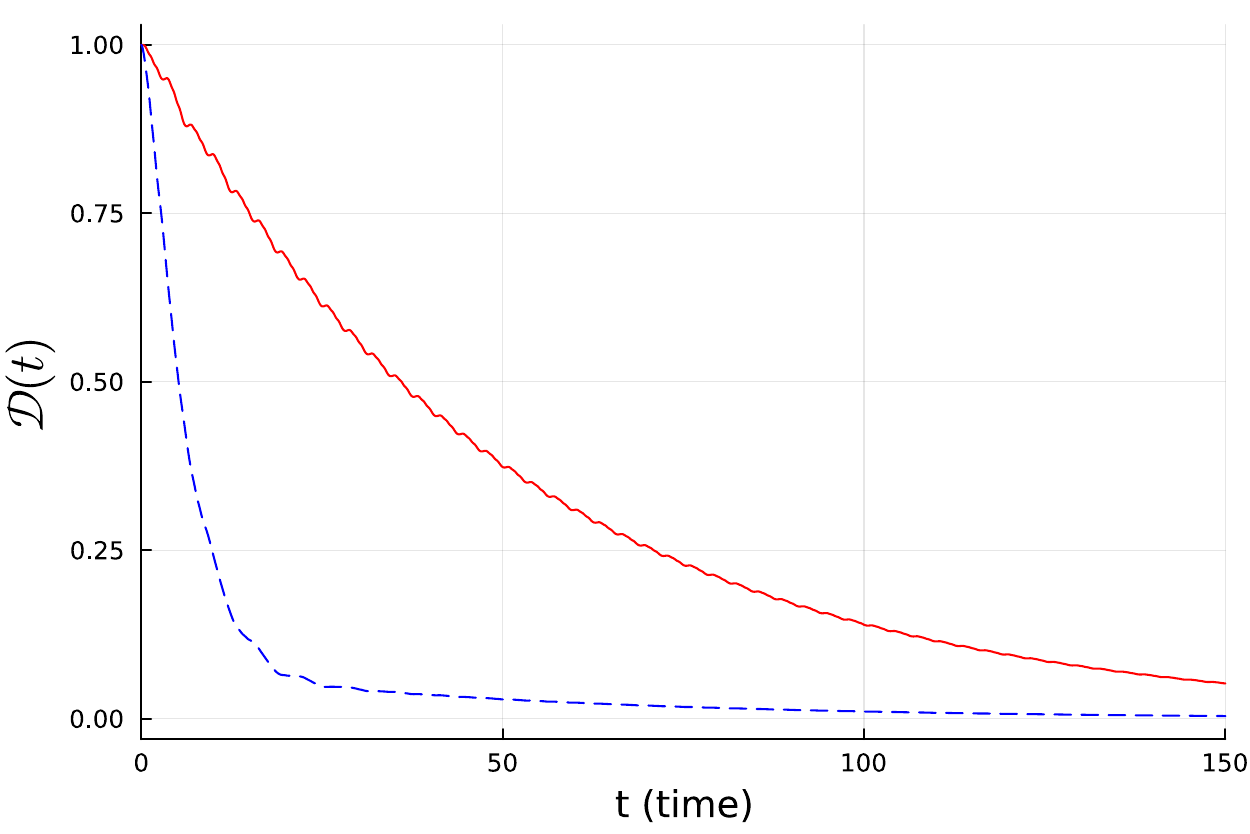}
         \includegraphics[width=\columnwidth]{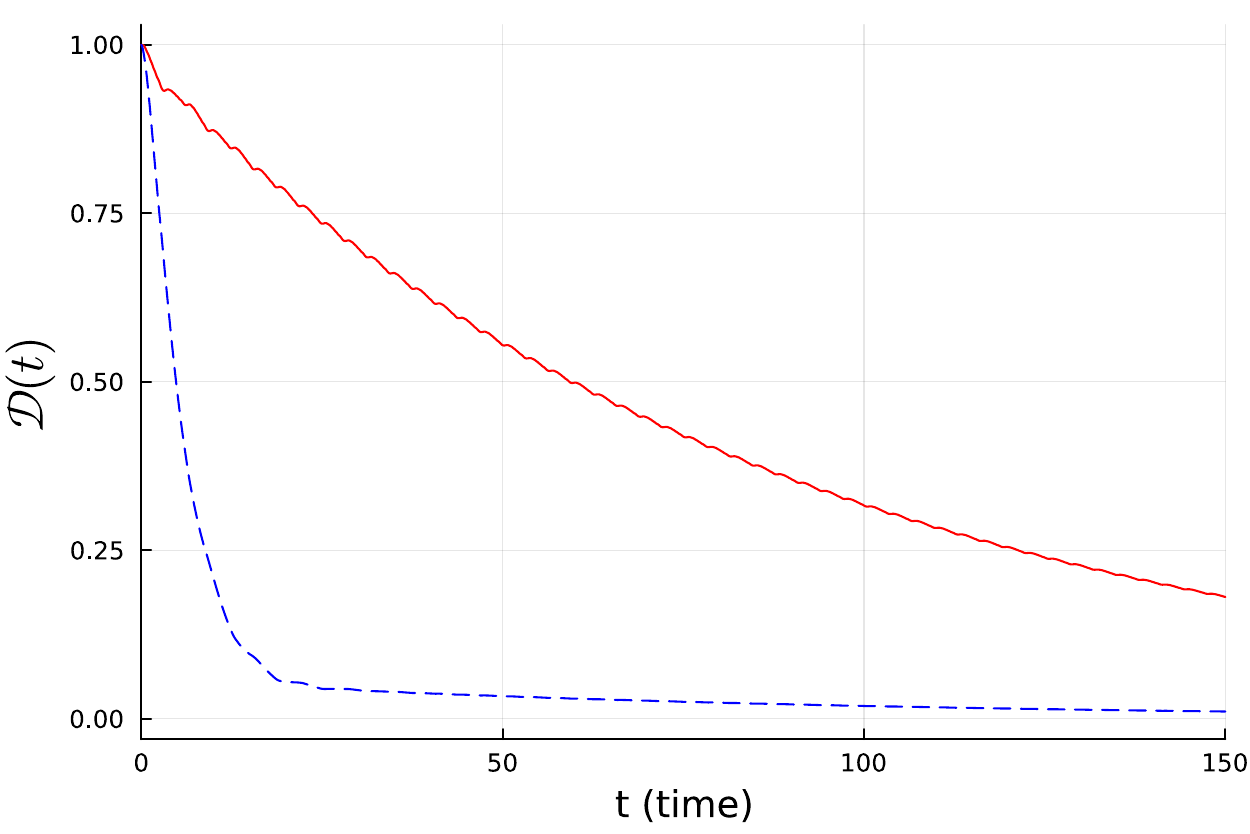}
         \caption{Time dependence of the trace distance at quasienergy crossings for two different values of the driving strength $\Omega$. Solid red line: initial pair maximizing non-Markovianity; dashed blue line: randomly drawn initial pair of states. Top and bottom panels correspond to $\Omega = 4.24\omega_0$ and $\Omega = 7.44\omega_0$, respectively.}
    \label{Fig:trace_dist_time}
\end{figure}

In Fig.~\ref{Fig:trace_dist_time}, we plot the behavior in time of the trace distance for two distinct values of the driving amplitude $\Omega$, both corresponding to quasienergy crossings. The trace distance is evaluated for both the pair of orthogonal states maximizing non-Markovianity and another randomly chosen pair of initial states. 
It can be observed that at quasienergy degeneracy information backflow processes are small but persistent throughout the entire time evolution. This behavior allows us to link non-Markovianity to longer coherence times: as the coherence time increases the non-Markovianity measure of Eq.~\eqref{eq:NM}, that is given by the sum of the heights of the revivals, grows.

\section{Numerical evaluation of $c^{n}_{i,j}$}
\label{eval_four_coeff}
\begin{figure*}
  \centering
  \begin{minipage}[t]{0.49\textwidth}
    \centering
    \includegraphics[width=\columnwidth]{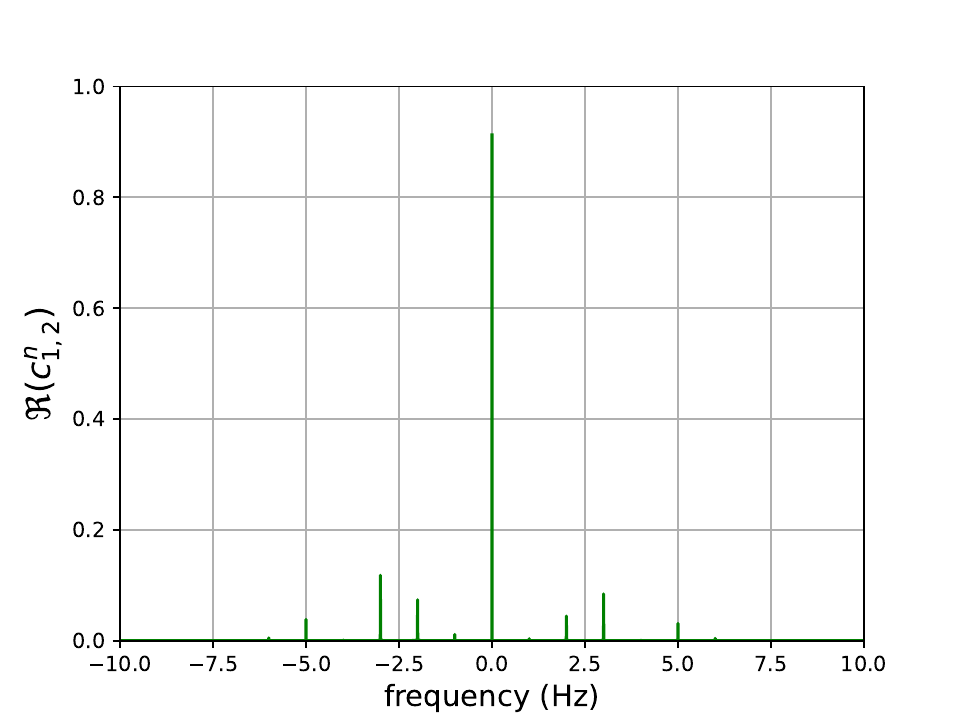}\par
    \includegraphics[width=\columnwidth]{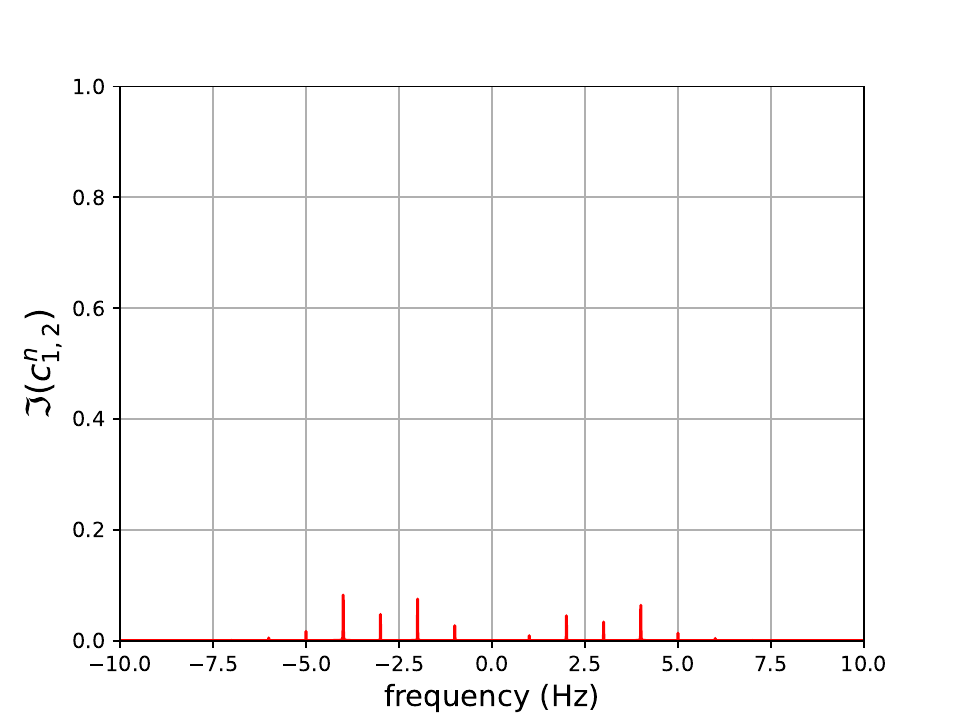}\par
    \includegraphics[width=\columnwidth]{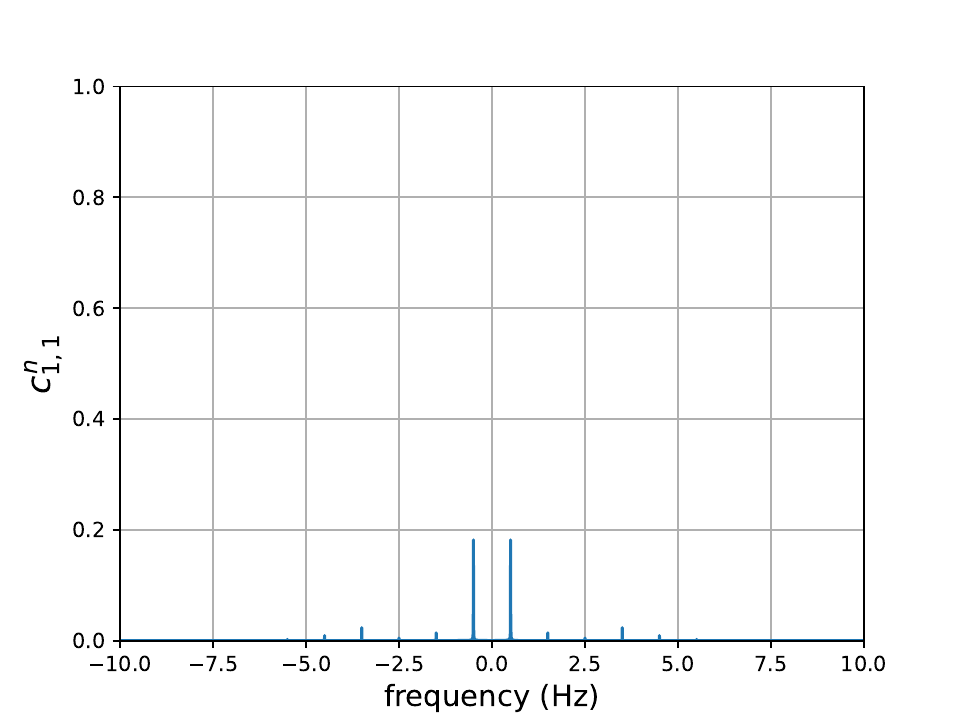}
    \begin{center}
    {(a) Degenerate case}    
    \end{center}
  \end{minipage}
  \hfill
  \begin{minipage}[t]{0.49\textwidth}
    \centering
    \includegraphics[width=\columnwidth]{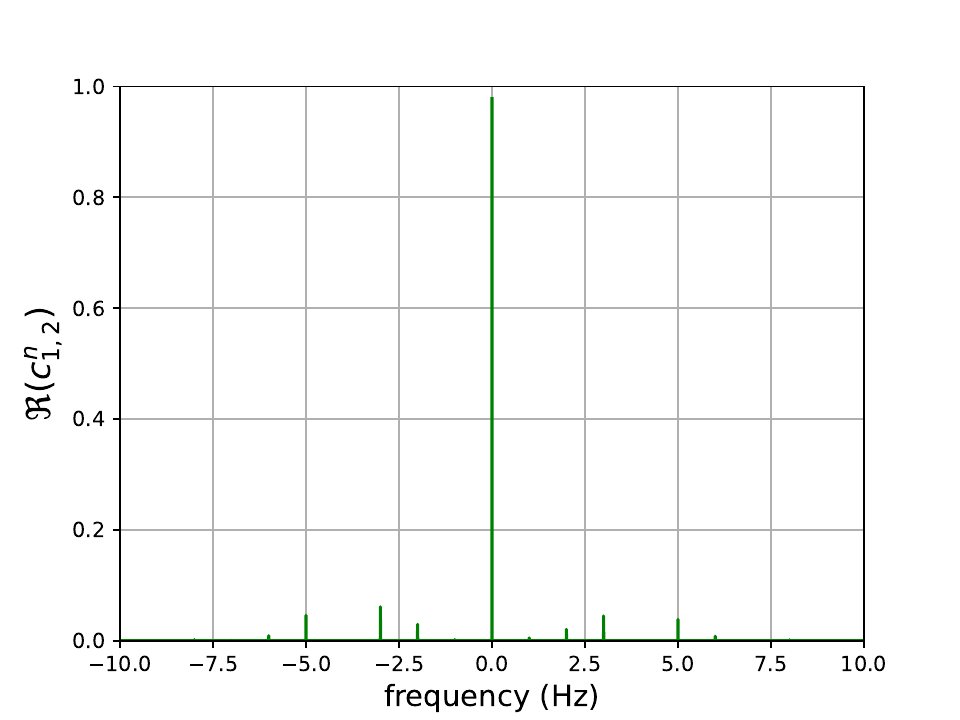}\par
    \includegraphics[width=\columnwidth]{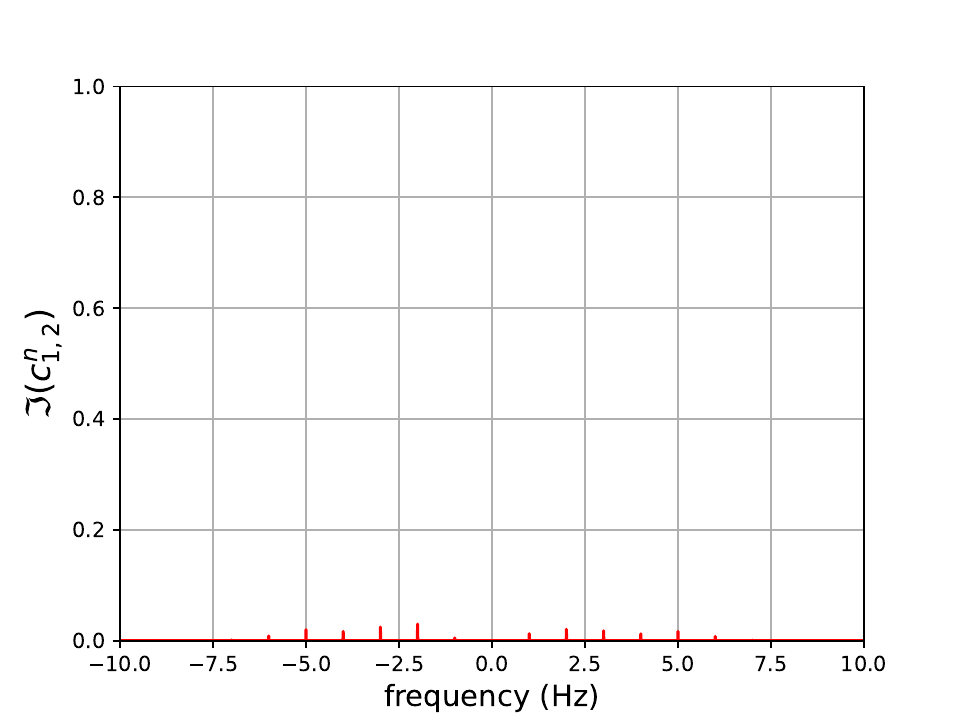}\par
    \includegraphics[width=\columnwidth]{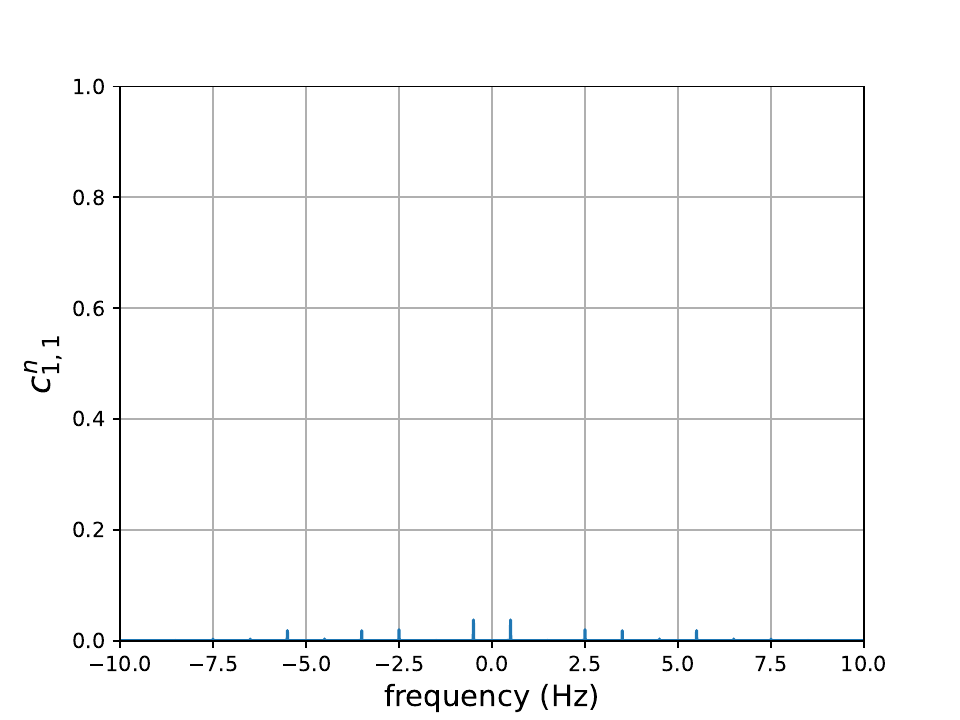}
    \begin{center}
    {(b) Non-degenerate case}    
    \end{center}
  \end{minipage}

  \caption{%
  Comparison between Fourier components $c^n_{ij}$ for degenerate and non-degenerate cases.
  Panel (a): coefficients $\Re(c^n_{12})$, $\Im(c^n_{12})$, and $c^n_{11}$ as a function of $n$ for
  $\Omega = 4.27\omega_0$, corresponding to a quasienergy crossing.
  Panel (b): same quantities for $\Omega = 6.5\omega_0$, away from the crossing.
  The imaginary part of $c^n_{12}$ is negligible for all $n$, while $c^n_{11}$ is appreciably
  different from zero only in the degenerate case for $n=\pm1$.}
  \label{fig:Degenerate_vs_NonDegenerate}
\end{figure*}

\begin{figure}[htb]
    \includegraphics[width=\columnwidth]{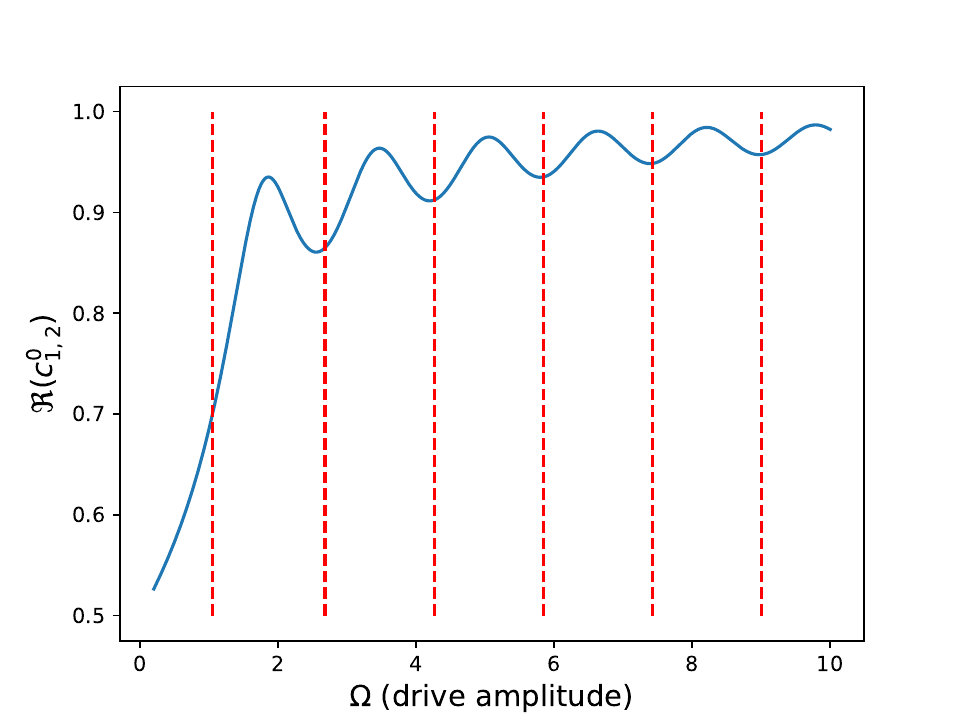}
    \includegraphics[width=\columnwidth]{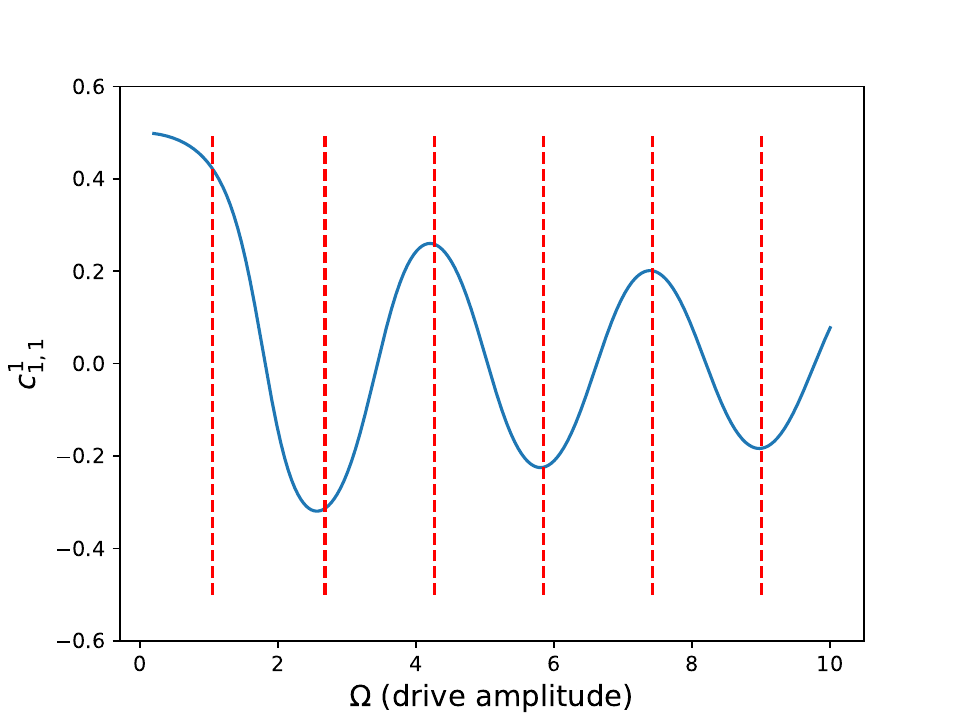}
    \includegraphics[width=\columnwidth]{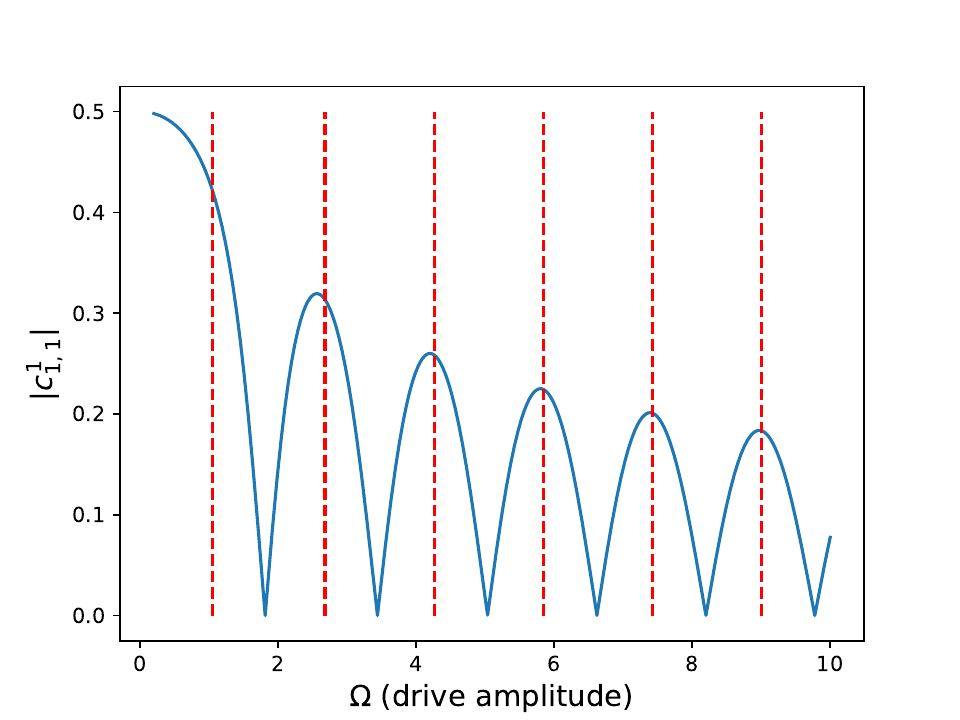}
  \caption{Fourier components $c^n_{ij}$ as a function of the driving amplitude $\Omega$. The dashed vertical lines denote the quasienergy crossings.}
  \label{fig:Sup_coeff_depend}
\end{figure}

The coefficients $c^{n}_{i,j}$ defined according to Eq.~\eqref{eq:four_coef} play a central role in determining the structure of the Floquet-Lindblad master equation and, consequently, the system's dynamics, since they determine according to Eq.~\eqref{eq:A_omega_f} the jump operators. In this work, we employ suitable approximations for these coefficients, based on their numerical evaluation, which is detailed in this Appendix. The coefficients are the Fourier components of the coupling operator in the Floquet basis, defined in Eq.~\eqref{eq:four_coef} which we report here for convenience
\begin{equation}
    c^n_{ij} = \int_0^T \frac{dt}{T} e^{-i n \omega_0 t} \bra{u_i(t)} \sigma_x \ket{u_j(t)}.
    \label{eq:SUP_four_coef}
\end{equation}
They can be computed numerically by standard means. We evaluate the time evolution of the Floquet basis states using the QuTiP library~\cite{Tanimura2020a,Johansson2013a,Julia_HEOM} and then implement a discrete Fourier transform algorithm.
The validity of the relationships $c^n_{1,1}=-c^n_{2,2}$ and $c^n_{1,2}=(c^n_{2,1})^*$ already presented in Sec.~\ref{sec_num_resu_theo_exp} leads to the evaluation of only two coefficients sets, namely $c^n_{1,1}$ and $c^n_{1,2}$, with $n\in \mathbb{Z}$.
For the sake of clarity we first  provide in Fig.~\ref{fig:Degenerate_vs_NonDegenerate} the Fourier coefficients for two fixed values of the driving amplitude $\Omega$ representative of the degenerate and non-degenerate cases. We then explicitly show in Fig.~\ref{fig:Sup_coeff_depend} 
the $\Omega$-dependency of the relevant coefficients. 

In Fig.~\ref{fig:Degenerate_vs_NonDegenerate}(a) the Fourier components $c^{n}_{1,2}$ evaluated at a crossing point of  the quasienergy (specifically $\Omega = 4.27\omega_0$) are shown. Similar results are obtained for all values of the driving strength corresponding to quasienergy crossings, as we shall see considering the dependence of the coefficients on the driving amplitude $\Omega$  in Fig.~\ref{fig:Sup_coeff_depend}. 
From this, we can deduce that the imaginary part of $c^{n}_{1,2}$ is negligible compared to its real part for all $n$. Moreover, the only real component significantly different from zero is obtained for $n = 0$, which can be approximated by 1, leading to:
\begin{equation}
c^{n}_{1,2} \approx \delta_{n,0}.
\label{sup_eq:aprox_1}
\end{equation}
On the other hand, the only relevant coefficients of the diagonal components correspond to $n = \pm 1$ and can be approximated by $0.2$, so that we have:
\begin{equation}
c^{n}_{1,1} \approx \frac{1}{5}(\delta_{n,-1} + \delta_{n,1}).
\label{sup_eq:approx_2}
\end{equation}
In Fig.~\ref{fig:Degenerate_vs_NonDegenerate}(b), the Fourier components are shown for the non-degenerate case (specifically $\Omega = 6.5\omega_0$). It is evident that approximation \eqref{sup_eq:aprox_1} still holds, while the approximation \eqref{sup_eq:approx_2} is no longer valid. In particular, all coefficients $c^{n}_{1,1}$ appear to be negligible in this regime. 

According to Eq.~\eqref{eq:A_omega_f} in the degenerate case we are  left with the Lindblad operators specified in Eq.~\eqref{jump_eff-deg}, while in the non-degenerate case we have to consider the operators Eq.~\eqref{jump_eff}, thus leading to two different master equations.

\begin{figure}[htb]
        \centering
        \includegraphics[width=\columnwidth]{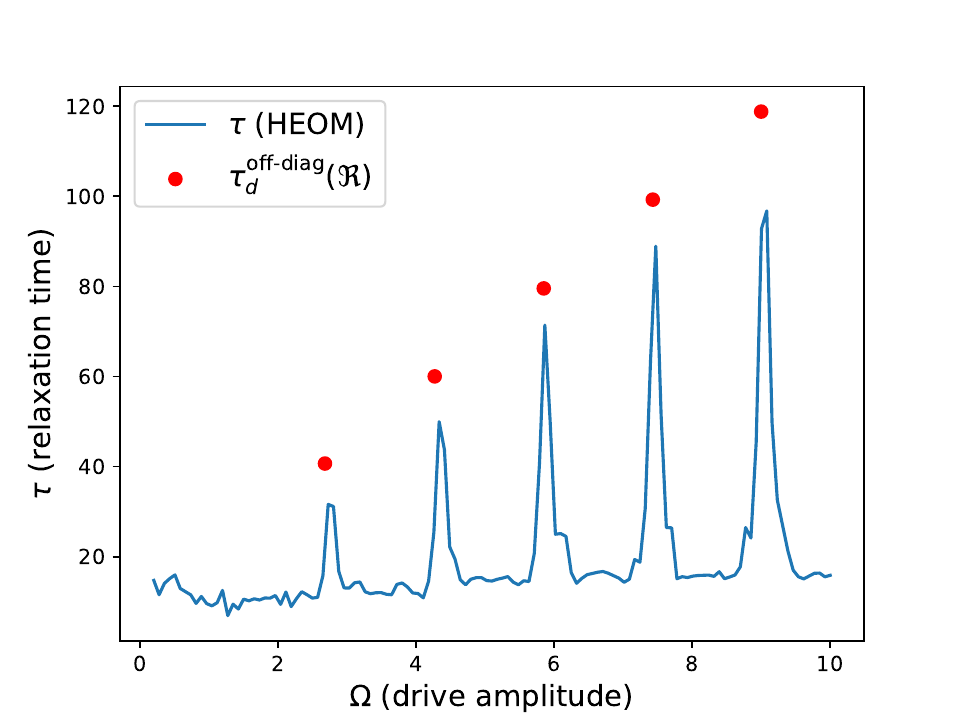}
        \caption{Comparison between relaxation times obtained by HEOM simulations and relaxation times for the real part of the coherences in correspondence of quasienergy crossings obtained by the analytical solution of the Floquet Lindblad master equation in the degenerate case.}
        \label{fig:tau_comparison}
    \end{figure}

These observations motivate a more detailed analysis of the dependence of $c^0_{1,2}$ and $c^1_{1,1}$ on the driving amplitude $\Omega$.
From Fig.~\ref{fig:Sup_coeff_depend} (top), it can be seen that approximation Eq.~\eqref{sup_eq:aprox_1} for the coefficient $c^0_{1,2}$ remains valid for all $\Omega \gtrsim 2.0\omega_0$, thereby confirming its general applicability. The behavior of $c^1_{1,1}$, shown in Fig.~\ref{fig:Sup_coeff_depend} (middle), is more intricate, as it exhibits sign changes. However, due to the explicit expression of the jump operators, only the relative phase between $c^n_{1,1}$ and $c^n_{2,2}$ is relevant for determining the Floquet-Lindblad master equation. Therefore, the sign change is irrelevant in light of the relation $c^1_{1,1} = -c^1_{2,2}$. Thus, for simplicity we can restrict the analysis to the absolute value $|c^1_{1,1}|$.
The plot of $|c^1_{1,1}|$ as a function of the driving amplitude in Fig.~\ref{fig:Sup_coeff_depend} (bottom) confirms the validity of approximation Eq.~\eqref{sup_eq:approx_2} at the quasienergy crossings. Importantly, as the driving amplitude $\Omega$ increases, the value of $|c^1_{1,1}|$ at the crossings slowly decreases. This trend plays a key role in explaining the secondary effect observed in Fig.~\ref{Fig:non_markovyrela_times}, namely the increase of the coherence time peaks with increasing $\Omega$. Indeed, using the numerically computed values of $c^1_{1,1}$ (see Fig.~\ref{fig:Sup_coeff_depend}), the relaxation time of the real part of the coherences at the quasienergy crossings can be accurately determined by inserting the numerical value into Eq.~\eqref{eq:larger}, instead of using the approximate value $|c^1_{1,1}| = \frac{1}{5}$, according to
\begin{equation}
    \tau_{\mathrm{D}}^{\mathrm{off-diag}}(\Re) =\frac{1}{2|c^{1}_{1,1}|^2\gamma_{z}}.
    \label{eq:tau_real}
\end{equation}
The relaxation times obtained from Eq.~\eqref{eq:tau_real} inserting the values used in the simulations for $\alpha$ and $\omega_c$ to evaluate $\gamma_z$ are plotted in Fig.~\ref{fig:tau_comparison} and compared with those extracted from the HEOM simulations.

The results show excellent agreement between the two quantities. In particular, the monotonic increase of the relaxation times with the driving amplitude $\Omega$ is accurately reproduced, further supporting the validity of our interpretation and its predictive power.

In the non-degenerate regime, $c^1_{1,1}$ varies significantly with $\Omega$, but it remains consistently smaller than $c^0_{1,2}$, so that this dependency does not influence our analysis. As a result, the dissipative channel associated with $\Sigma_z$ always has the smallest rate. For simplicity, we have approximated the absolute value of $c^1_{1,1}$ by its average value over the considered values of the driving strength, leading to $\bar{c}^1_{1,1}  \approx 0.2$. 

\section{Dependence on cutoff and driving frequency}
\label{Appendix_other_params}
\begin{figure*}[htb]
  \includegraphics[width=\columnwidth]{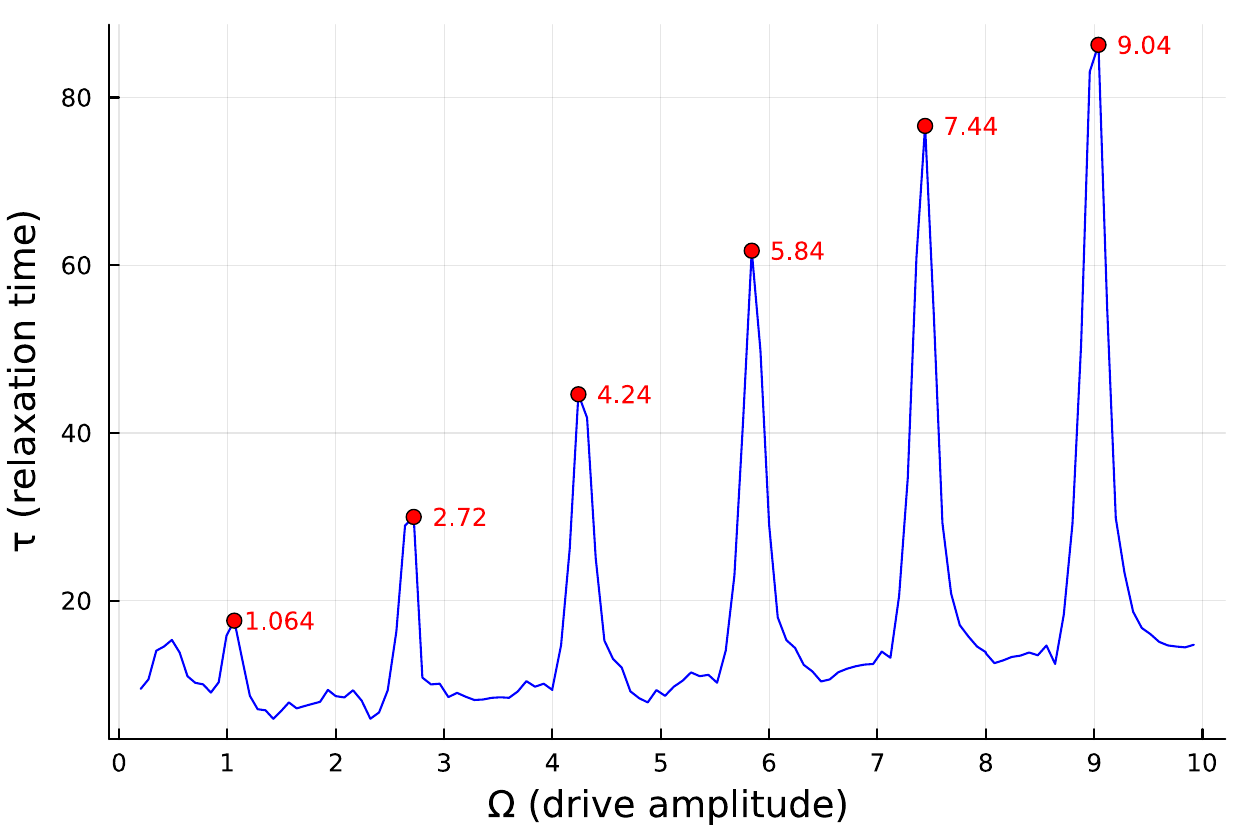}
  \includegraphics[width=\columnwidth]{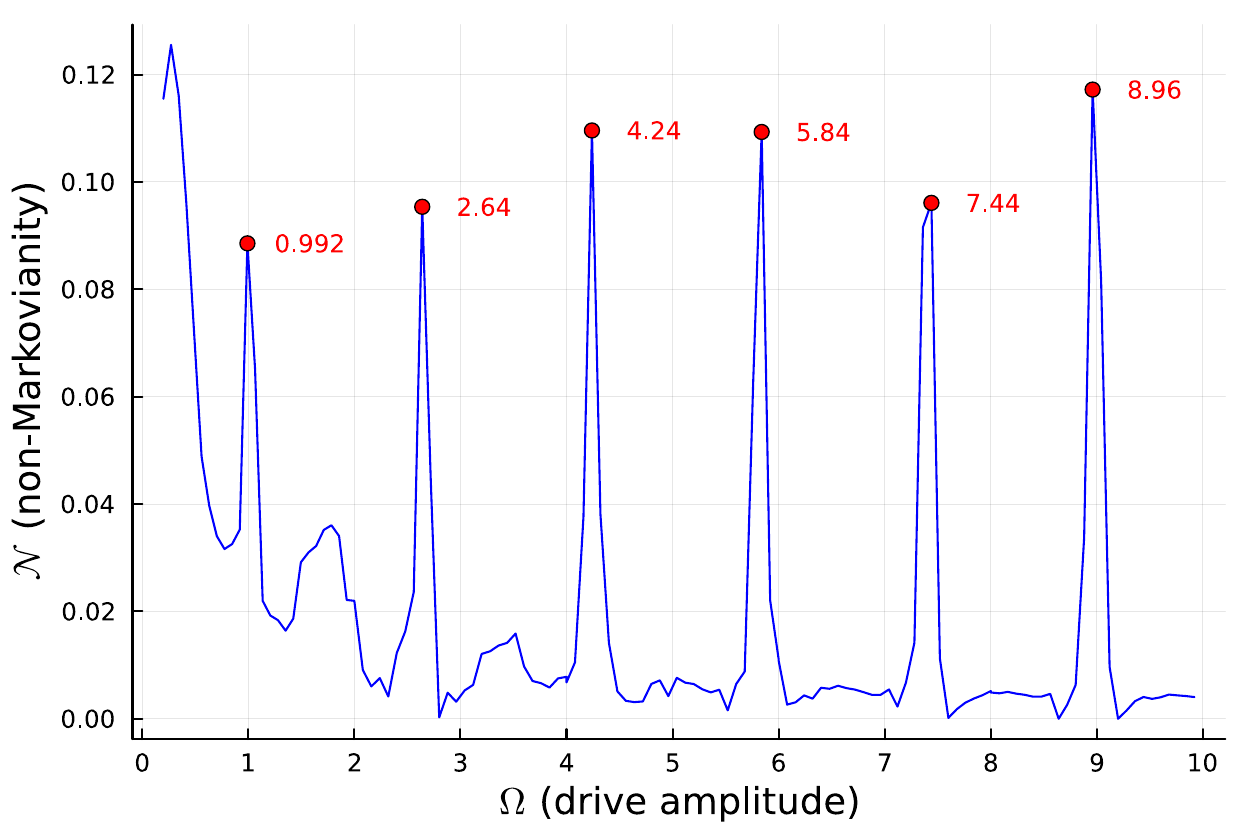}
  \caption{Relaxation time (left) and non-Markovianity (right) for the driven spin-boson system with reduced cutoff frequency $\omega_c = \omega_0/2$. Red dots denote the maxima positions.
  The driving frequency is set to the resonant value $\omega = \omega_0$, so that quasienergy crossings occur at
  $\Omega = \{1.05, 2.68, 4.27, 5.85, 7.43, 9.00\}\omega_0$.}
  \label{Fig:cut_05}
\end{figure*}

We emphasize the robustness of the emergence of these peaks in the
relaxation times, and hence of non-Markovianity, at quasienergy
crossings, by presenting numerical simulations in different parameter regimes.
In the simulations presented in Sec.~\ref{sec_num_resu_theo_exp}, the cutoff and driving frequencies are chosen to be equal to the two-level system energy splitting, i.e., $\omega_c = \omega = \omega_0$.
Here, we consider the behavior of relaxation times and non-Markovianity for a different cutoff frequency.
In Fig.~\ref{Fig:cut_05} we present the numerical results of relaxation times  and non-Markovianity  for $\omega_c =\omega_0/2$. The peak-like structure in both the relaxation times and non-Markovianity, as well as the one-to-one correspondence between the peaks, is immediately seen to be preserved by modifying the cutoff frequency.

\begin{figure*}[htb]
    \includegraphics[width=\columnwidth]{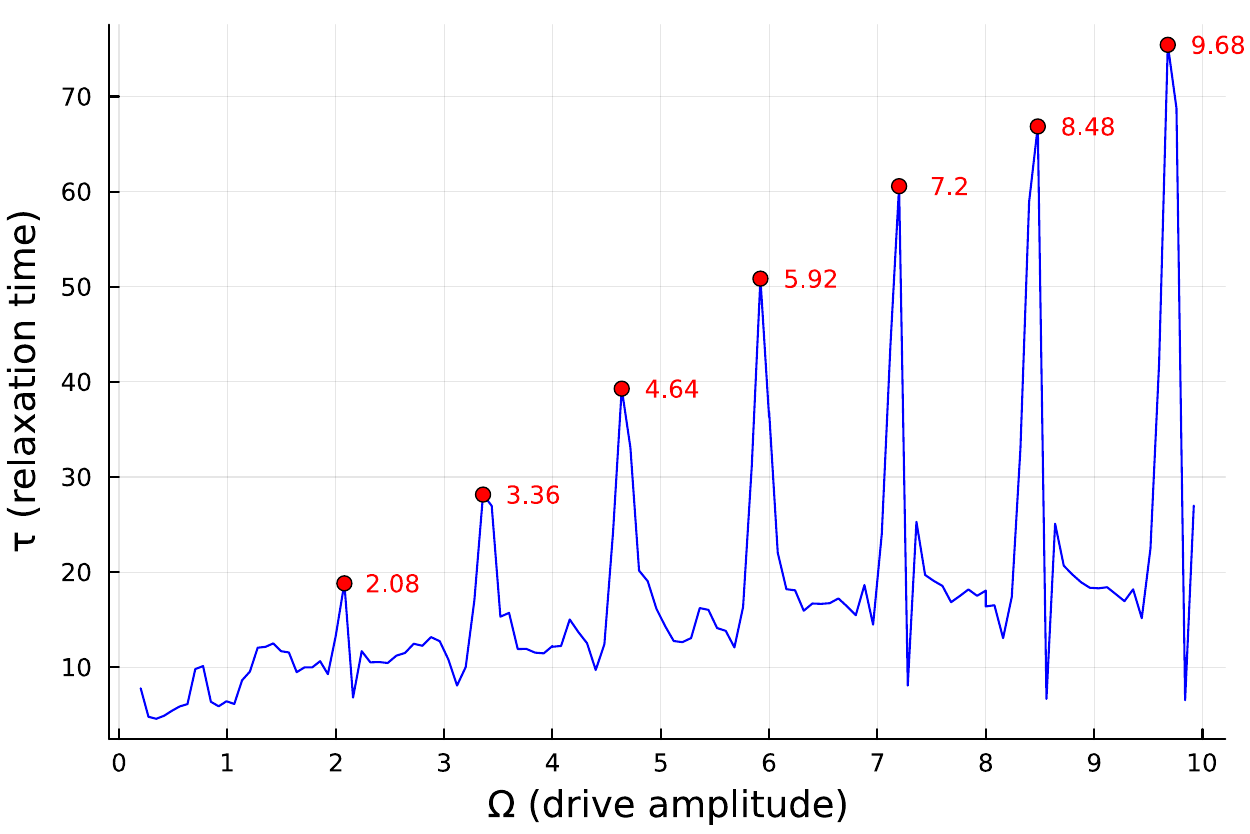}
  \includegraphics[width=\columnwidth]{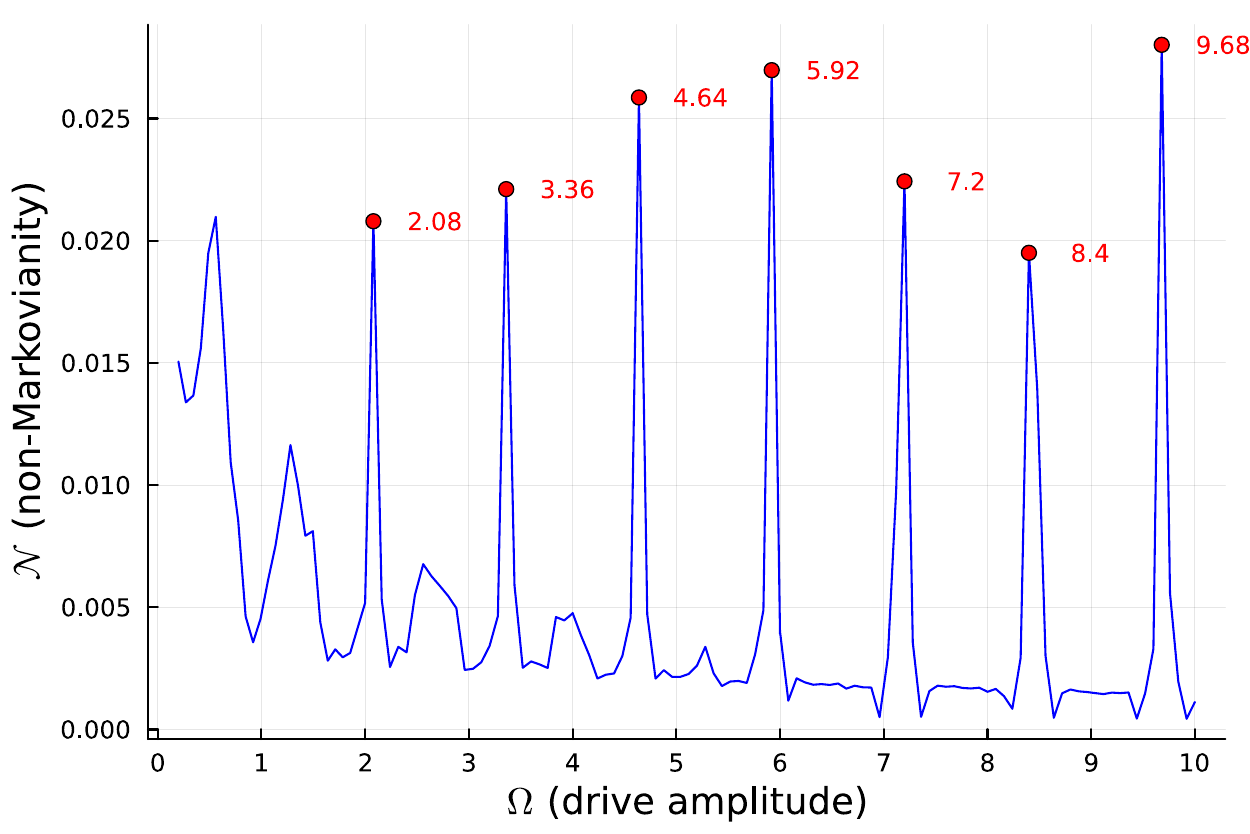}
  \par  
  \includegraphics[width=\columnwidth]{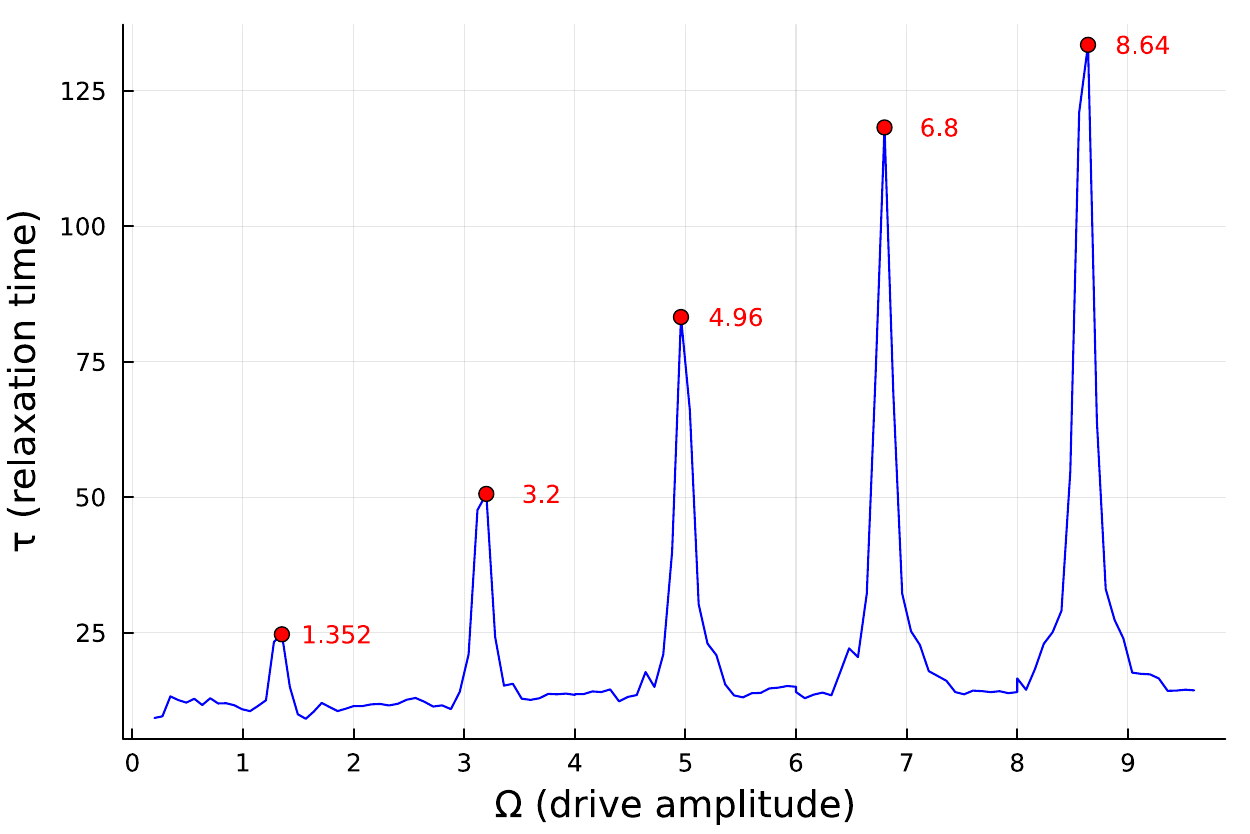}
  \includegraphics[width=\columnwidth]{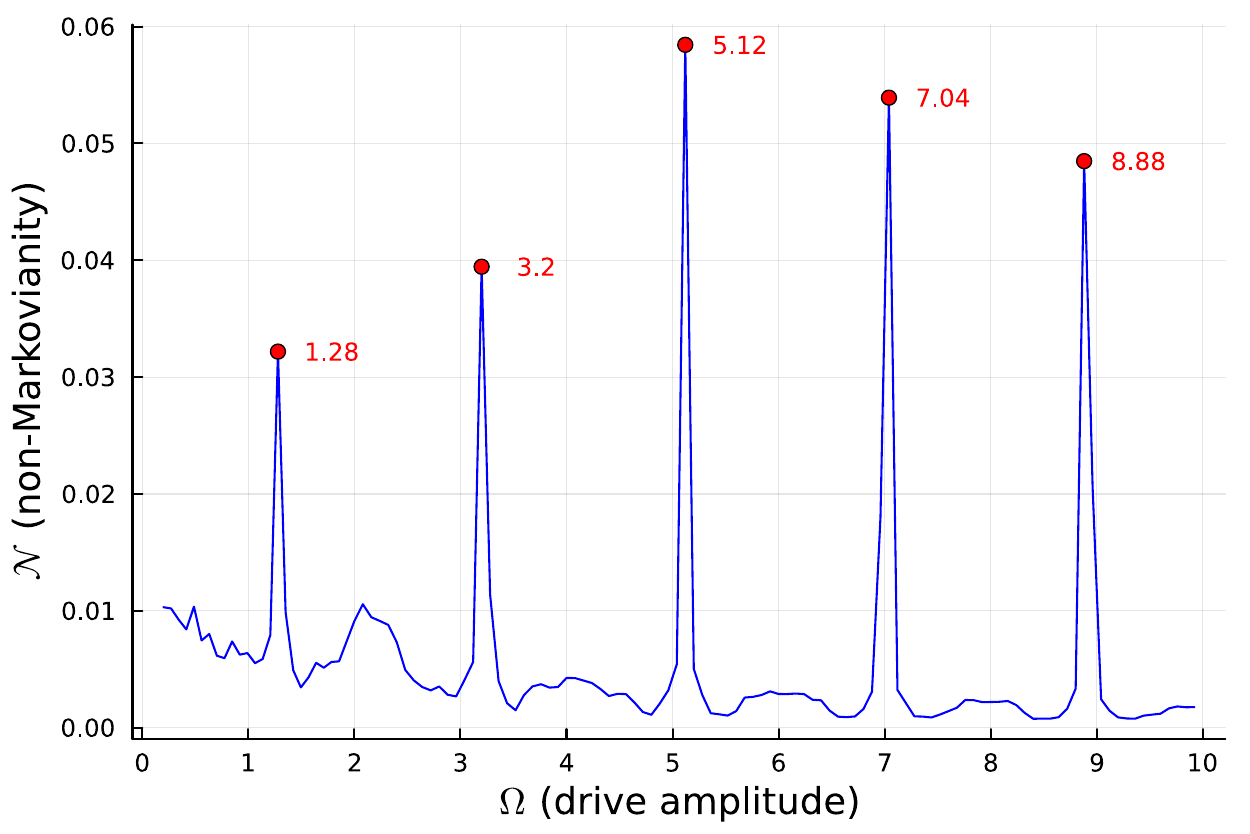}
  \caption{%
  Relaxation time (left) and non-Markovianity (right) for the driven spin system
  with off-resonant driving frequency $\omega = 0.8\omega_0$ (top) and $\omega = 1.2\omega_0$ (right).
  Red dots indicate the maxima corresponding to quasienergy crossings at
  $\Omega = \{0.76, 2.10, 3.39, 4.66, 5.92, 7.19, 8.45, 9.71\}\omega_0$ (for $\omega = 0.8\omega_0$) and $\Omega = \{1.32, 3.24, 5.14, 7.04, 8.93\}\omega_0$ (for $\omega = 1.2\omega_0$).}
  \label{Fig:off_08_12}
\end{figure*}

Furthermore, we consider the situation in which the driving is not resonant.
Indeed, the mechanism considered in this work takes place at the quasienergy crossings that appear for any driving frequency.
In Fig.~\ref{Fig:off_08_12} we present the numerical results for relaxation times (left panels) and non-Markovianity (right panels) in both the red detuned and blue detuned cases, $\omega = 0.8\omega_0$ and $\omega = 1.2\omega_0$ respectively. 
It can be observed that the peak structure is maintained even under the two off-resonant conditions. Moreover, the exact correspondence between quasienergy crossings and the peaks in non-Markovianity and relaxation times remains preserved.

\clearpage
\nocite{sup_data}
\bibliography{biblio,ref}

\end{document}